\documentclass[usenatbib,usegraphicx]{mn2e}

\title[V2494 Cyg: A unique FU~Ori type object]{V2494 Cyg: A unique FU~Ori type object in the Cygnus OB7 complex}
\author[Tigran Yu. Magakian et al.]{Tigran Yu. Magakian,$^{1}$\thanks{E-mail:
tigmag@sci.am}\thanks{Some results, discussed in this
paper, were presented briefly in JENAM-2007 conference and IAU 243
symposium. } 
Elena H. Nikogossian,$^{1}$
Tigran Movsessian,$^{1}$
Alexei Moiseev,$^{2}$ 
\newauthor Colin Aspin,$^{3}$
Chris J. Davis,$^{4}$
Tae-Soo Pyo,$^{5}$
Tigran Khanzadyan,$^{6}$
Dirk Froebrich,$^{7}$
\newauthor Michael D. Smith,$^{7}$ 
Gerald H. Moriarty-Schieven,$^{8}$ 
Tracy L. Beck$^{9}$\\
$^{1}$V.A. Ambartsumyan Byurakan Astrophysical Observatory,
  0213 Aragatsotn reg., Armenia\\
$^{2}$Special Astrophysical Observatory, RAS, 369167,
  N.Arkhyz, Zelenchukskij r., Karachaj-Cherkessian Rep., Russia \\
$^{3}$Institute for Astronomy, University of Hawaii, 640
  North A`oh\=ok\=u Place, Hilo, HI, 96720, USA \\
$^{4}$Astrophysics Research Institute, Liverpool John Moores University,
Egerton Wharf, Birkenhead,Wirral, CH41 1LD, UK\\
$^{5}$Subaru Telescope, National Astronomical Observatory
    of Japan, 650 North A`oh\=ok\=u Place, Hilo, HI 96720, USA
   \\
$^{6}$Max-Plank-Institut fuer Radioastronomie, Auf dem Huegel 69, D-53121 Bonn, Germany\\
$^{7}$Centre for Astrophysics \& Planetary Science, School of Physical Sciences, The University of Kent, Canterbury CT2 7NH, England\\
$^{8}$Herzberg Institute of Astrophysics 5071 West Saanich Road Victoria, British Columbia, Canada V9E 2E7\\
$^{9}$Space Telescope Science Institute, 3700 San Martin Drive, Baltimore, MD 21218, USA
    }
\begin{document}

\date{Accepted .. Received ..; In original form...}

\pagerange{\pageref{firstpage}--\pageref{lastpage}} \pubyear{2002}

\maketitle

\label{firstpage}

\begin{abstract}
A photometric and spectral study of the variable star
V2494~Cyg in the L~1003 dark cloud is presented. The
brightness of the star, formerly known as HH~381~IRS, increased by 2.5 mag in R (probably in the
1980s) and since then has remained nearly constant. Since the
brightness increase, V2494 Cyg has illuminated a bipolar cometary
nebula. The stellar spectrum has several features typical of the FU
Ori type, plus it exhibits very strong H$\alpha$ and forbidden
emission lines with high-velocity components. These emission lines
originate in the HH jet near the star. The kinematic age of the jet is
consistent with it forming at the time of the outburst leading to the
luminosity increase.  V2494 Cyg also produces a rather extended
outflow; it is the first known FUor with both an observed outburst and
a parsec-sized HH flow. The 
nebula, illuminated by V2494 Cyg, possesses similar morphological
and spectral characteristics to Hubble's Variable Nebula (R Monocerotis/NGC 2261).
\end{abstract}

\section{Introduction}

Objects classified as FU~Ori type (FUors), though very rare, provide clues to
the understanding of star formation processes and the evolution of
pre-main sequence (PMS) objects. Their main features are well-known
\citep{herbig2,hartmannkenyon}.  Their key attribute is a sudden
significant rise in brightness (up to 5 -- 6 magnitudes in V) in a
relatively short period (0.5 -- 10 years), after which the object
usually remains bright with little variability for tens of
years. Other typical qualities include:

\begin{description}
\item{a close association with a compact reflection nebulae;} 
\item{a large excess of radiation at UV and IR wavelengths,}
\item{a spectral type after the eruption similar to F--G supergiants, with}
\item{earlier type spectra in the UV, and later type in the IR,}
\item{deep CO absorption features,}
\item{the presence of prominent P~Cyg profiles in H$\alpha$ and Na~D
  lines, indicating mass loss at up to 1000 km s$^{-1}$ velocity, and}
\item{the presence of Li~I spectral lines, indicating the youth of the
  star.}  
\end{description}

Only about twenty such objects are known at present
\citep{reipurthaspin3}, half of them being classified as
``classical'' FUors (i.e. those where the outburst has been observed),
and the other half, classified as FUor-like, are represented by stars
which have similar spectral characteristics but where no outbursts
have been detected (i.e. it is assumed that they were discovered after
the outburst).

According to the generally accepted model of the FUor phenomenon, the
sudden brightening is probably due to a significant increase in
accretion (up to 10$^{-4}$ M$_{\sun}$) from the circumstellar matter
onto the T~Tau star \citep{hartmannkenyon}. Disturbances to the
accreting disk creating such an increase may be caused by thermal
instabilities \citep{belllin}, or may be due to a close companion
\citep{bonnellbastien,reipurthaspin2} or giant planet in the disk
\citep{lodatoclarke}. A comprehensive review of these models can be
found in the book of \cite{hartmann2009}. However,
notwithstanding the successful explanation of some features of FUors
by this approach, many questions still remain
open. \citet*{herbigetal}, for example, suggest that a number of the
observational aspects may be better explained if the star itself,
rather than an accretion disk, is responsible for the FUor flare-up.
For example, such an increase in brightness can occur for stars with
anomalous high rotation velocities \citep{larson}.
One should
also keep in mind that there exist a number of PMS stars
which display certain FU Ori features (e.g. V1331~Cyg, V1647~Ori) and are considered by
some authors to be pre-FUors or post-FUors. Besides, yet another and also not numerous class  of PMS variables, EXors, can be related to the FU Ori phenomenon \citep[see][and references therein]{reipurthaspin3}.  Thus, in future, when the volume of observational data increases, the criteria for defining the object as FUor may be expanded. In this paper, however, we will keep close to the genuine FU~Ori picture. 

No FUor has so far had more than one outburst detected. However, from
statistical considerations, it is usually accepted that FUors should
exhibit recurring eruptions with a timescale of order $10^{4}$ years
\citep*{herbig2,hartmannkenyon,scholzetal}.  In this connection it is tempting to
make a direct connection between probably-repetitive FU~Ori outbursts
and extended Herbig-Haro (HH) outflows with multiple working surfaces
(implying the periodic release of of matter from young stellar
sources)
\citep{reipurth,reipurthaspin1,reipurthbally,herbigetal,movsessianetal2}.

Another crucial but still open question concerns the fraction of
typical low-mass stars which undergo the FU~Ori phenomenon in their early
evolution. To date, only
classical T Tauri stars have been directly found to undergo FU Or outbursts. For a long time the only object with a known pre-outburst spectrum remained the V1057~Cyg star; the recent discovery of a FU~Ori type eruption in LkH$\alpha$~188-G4 \citep{milleretal}  raises the number of such objects to two. Both stars prior to eruption were  classical T~Tau stars.

However, the timescale between outbursts of order 10,000 years makes these
events extremely rare. If all accreting young stars undergo the outbursts, then it is probable that the first discoveries will be within the largest and easiest observed class of
accreting object. 
If this timescale applies to other types of YSO, then similar outbursts may be revealed in them once similar extended surveys are performed.

 The situation, doubtless, will be elucidated in the future, when we will have pre-outburst spectral energy distributions for newly discovered FUors,    which would then probably occur in the  
large number of the presently performed infrared surveys such as 2MASS, IRAS, Spitzer, AKARI and
WISE. The work of \citet{scholzetal} should be mentioned as an example of ongoing surveys for FU~Ori objects.

We present here the results of new observations in the optical
range of the probable FU~Ori-like object V2494~Cyg, also known as
IRAS~20568+5217 and HH~381~IRS \citep{reipurthaspin1}. This object is 
associated with
a bright cometary nebula and several HH objects \citep*{devineetal}. It
is located in the dark cloud L~1003, belonging to the Cyg~OB7
star-forming complex \citep{movsessianetal1} where another FU~Ori
type object, V2495~Cyg, has been identified
\citep{movsessianetal2,magakianetal2010}.  V2494~Cyg possesses a jet
and a large bipolar outflow traced by HH objects and Molecular
Hydrogen emission-line Objects (MHOs). These include HH~382 and HH~966
as well as MHO 900, 901, 902 and 904 \citep{magakianetal2010,
  khanzadyanetal}. 

The similarity of the infrared spectrum of V2494~Cyg to
other FUors has already been noted by \citet{reipurthaspin1}. Further
studies in the infrared range have confirmed this conclusion
\citep{aspinetal}. At the same time its recent brightening in the
optical range was also revealed \citep{magakianetal1}.

The distance to the L~1003 cloud was estimated only roughly.
Hereafter we will assume the value 800 pc for V2494~Cyg and for the
entire Cyg~OB7 complex \citep*{dobashietal1}, consistent with all
our previous works
\citep{aspinetal,magakianetal2010,aspinetal2011,khanzadyanetal}.

In Sect. 2 we describe the observational methods and data reduction.
In Sect. 3 results of the photometry and the spectroscopy are
presented. In Sect. 4 we discuss separately the central star and
outflow and summarize our conclusions. 
\section{Observations and data reduction}

Our observations of V2494~Cyg include imaging and photometry as well
as spectroscopy in various modes. The observing logs are summarized in
Tables \,\ref{RIphot} and \,\ref{specobs}. We first
describe these observations in more detail.

\subsection{Imaging}

Image archives as well as new observations were used for photometry
of the star and for analysis of the morphology of the associated cometary nebula. Archival
data were from the DSS-1, Quick-V, DSS-2 and SuperCOSMOS (SSS) digital
sky atlases, as well as images
from the IPHAS survey, obtained on 2003 November 10-14
\citep{drewetal}. In addition, a number of photographic plates
from the Tautenburg plate archive, obtained in 1975-1985, were checked
for the presence of the object (Helmut Meusinger, private
communication).

New imaging data, to check the recent variability of the object, were
obtained on 2006 September 22-24 in R and I using the Subaru Prime
Focus Camera \citep[Suprime-Cam,][]{miyazakietal} mounted at the prime
focus of the 8.2-m Subaru Telescope atop Mauna Kea, Hawaii (described
in the previous work: \citealt{magakianetal2010}); two direct images,
obtained in 2008 with a CCD-camera on the ZTSh 2.6-m telescope of the Crimean
Observatory; plus a number of images, obtained with exposures of 600
sec with the ByuFOSC-2 \citep{movsessianetal} and SCORPIO
\citep{afanasievmoiseev} prime focus cameras on the 2.6-m telescope
ZTA of Byurakan Observatory in 2008-2010.

The Byurakan and Crimean images were reduced with standard procedures
for aperture photometry.  Magnitudes were converted to the Cousins
photometric system. For photometric calibration, standard stars from
the NGC~7790 cluster \citep{landolt} were taken. In addition, we used
three non-variable stars in the L~1003 cloud as secondary standards,
for which we determined R$_{C}$ and I$_{C}$ magnitudes.  To compare
the new and archive data and to avoid systematic errors as far as
possible, we re-measured all sky survey images, calibrating them by
nearby stars, magnitudes of which were taken from the GSC 2.3.2
catalogue which could be considered as the best photometric source for
these purposes \citep*{mickaelianetal}.

The observation dates of all images used in this paper are listed in
Table~\ref{RIphot} along with the photometric results.

\subsection{Slit spectroscopy}

The optical spectrum of the star V2494~Cyg and of the its associated nebula was observed in 2007 January 10 with the
6-m telescope of SAO (Russia) using the SCORPIO multi-mode focal reducer
\citep{afanasievmoiseev}, mounted at the prime focus of the
telescope. In the long-slit mode the instrument was equipped with the
VPH1800R grating, which provides a $\sim$2.5 \AA\ resolution and a
spectral range of 6100-7100 \AA.  A 2048 $\times$ 2048 EEV 42-40 CCD
was used for the detector; after binning by 1 $\times$ 2 the scale was
0.36 arcsec per pixel along the slit. Total exposure time was 1200
sec and seeing was 1.6 arcsec. The position angle of the slit during
the observations was 343$\degr$; this direction is aligned with the
jet and the axis of the reflection nebula.

Data reduction was performed with a package of procedures developed in
IDL by A. Moiseev at SAO.  This package includes standard operations,
such as flat-fielding, wavelength and flux calibration. Further
analysis was done with the ESO MIDAS image processing system.

We also made an attempt to obtain a long-slit spectrum of the knot A
in HH~382 group, using the SCORPIO  focal reducer on the 2.6-m
telescope in Byurakan.  The measurable spectrum of HH~382A was
registered in the red range ($\lambda\lambda5900-6900$) on 2008 June
13, with a total exposure time of 3600 sec and 0.5 \AA/pix dispersion.

\subsection{Imaging Spectroscopy}

Imaging spectroscopy observations were obtained using the SCORPIO
prime focus multi-mode device equipped with a  scanning Fabry-Perot etalon
and 2048 $\times$ 2048 EEV 42-40 CCD, mounted on the 6-m SAO
telescope.  The field around V2494~Cyg, which includes the jet and
nearby HH knots, was observed in the H$\alpha$ line on the 28th of
June 2008.  Observations were performed with 4 $\times$ 4 pixel
binning to reduce the readout time and enhance the S/N ratio;
512 $\times$ 512 pixel images were obtained for each spectral
channel. The field of view was 6.1 $\times$ 6.1 arcmin with a
scale of 0.71 arcsec per pixel. An interference filter with FWHM = 15
\AA\ centered on the H$\alpha$ line was used for
pre-monochromatization.  For our observations we used a Queensgate
ET-50 etalon operating in the 501st order of interference at the
H$\alpha$ wavelength, and providing an instrumental profile
of FWHM $\approx$ 0.8 \AA\ (or $\sim$ 36 km s$^{-1}$) for a range of
$\Delta\lambda$=13.2\AA\ (or $\sim$ 605~km\,s$^{-1}$), free from order
overlapping. The number of spectral channels was 36 and the size of a
single channel was $\Delta\lambda\approx$ 0.37 \AA\ ($\sim$
17~km\,s$^{-1}$).

We reduced our Fabry-Perot  observations using software developed
at SAO \citep{moiseevegorov}. After primary data reduction,
subtraction of night sky lines and wavelength calibration, the
observational material represents data cubes where each point in
the 512 $\times$ 512 pixel field contains a 36 channel spectrum. For
data analysis we used the ADHOCw software, developed by the
Interferometry Group of Marseille
Observatory\footnote{http://www.oamp.fr/adhoc/adhocw.htm} (see
\citealt{garridoetal} for an example of Fabry-Perot  data reduction).

\section{Results}
\subsection{The nebula imaging}

As already mentioned above, this spectacular bipolar nebula is missing
in common catalogs.  It was described for the first time in the work of
\cite{devineetal}.  We have reviewed its photometric history and found
that the object became significantly brighter in the period between
the DSS-1 and DSS-2 surveys, i.e. from 1952 to 1989.  In 1952 the
central star, which is associated with the IRAS~20568+5217 source, is
detectable only on the DSS-1 red image, near the plate limit. In
1983, in the Quick-V sky survey, the star is definitely brighter
(especially when one considers the different bandwidths
of both surveys and the very red colour of the star) and traces of the
nebula can be seen. On the DSS-2 R (1990) and I (1991) charts the
nebula is well developed and is also now visible in the bluer
wavebands (DSS-2 B, 1989). The discovery image, obtained in
1995-1996 \citep{devineetal}, shows the nebula with essentially
the same appearance as in our new images.  Thus, the increase of the
brightness started after 1952 and, probably, not long before 1983.

\begin{figure*}
\includegraphics[width=30pc]{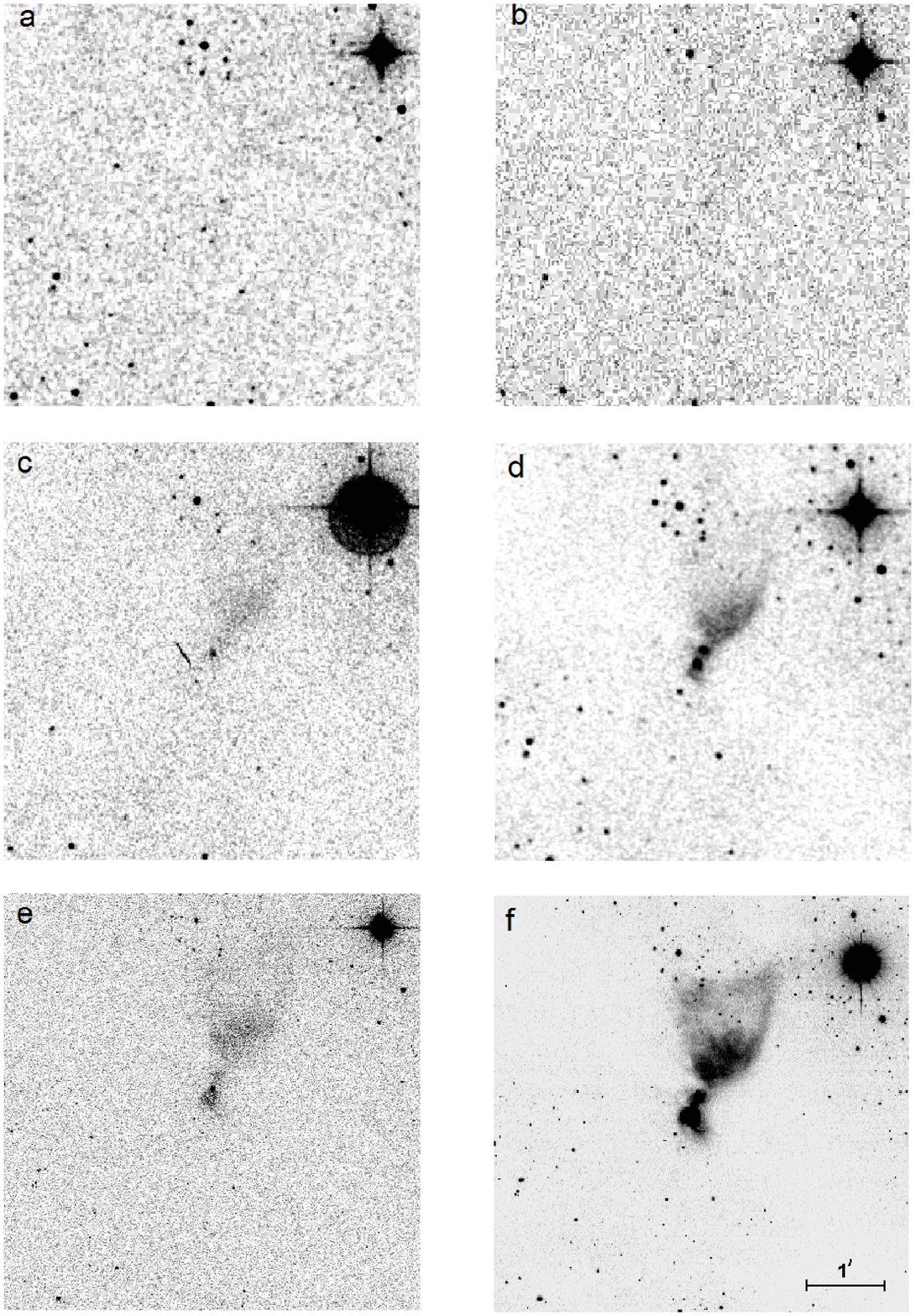}
\caption{The appearance of the V2494 Cyg and its associated nebula in
  various epochs. Panels: a) - POSS-1 R, 1952 (SSS survey); b) -
  Quick V, 1983; c) DSS-2 B, 1989; d) - DSS-2 R, 1991; e) IPHAS
  survey, R, 2003; f) - Subaru R, 2006. North is up, east is left.
  The object is located in the centre of the panels. The streak to the east
  from the star on the panel c) is an artifact
  on DSS-2. Note the bright nebulous knot in NW direction from the
  star on the panel d). The size of the each panel is 
  about 5.3 $\times$ 5.3 arcmin.}
\label{IRASneb}
\end{figure*}

For the last 15-20 years the star seems to have reached the light
curve plateau. Its photometric history is described
in the next section. We show images of the object taken at 
various epochs in Fig.\,\ref{IRASneb}.

\begin{figure*}
\begin{minipage}{0.49\linewidth}
\centering
\includegraphics[width=0.85\linewidth]{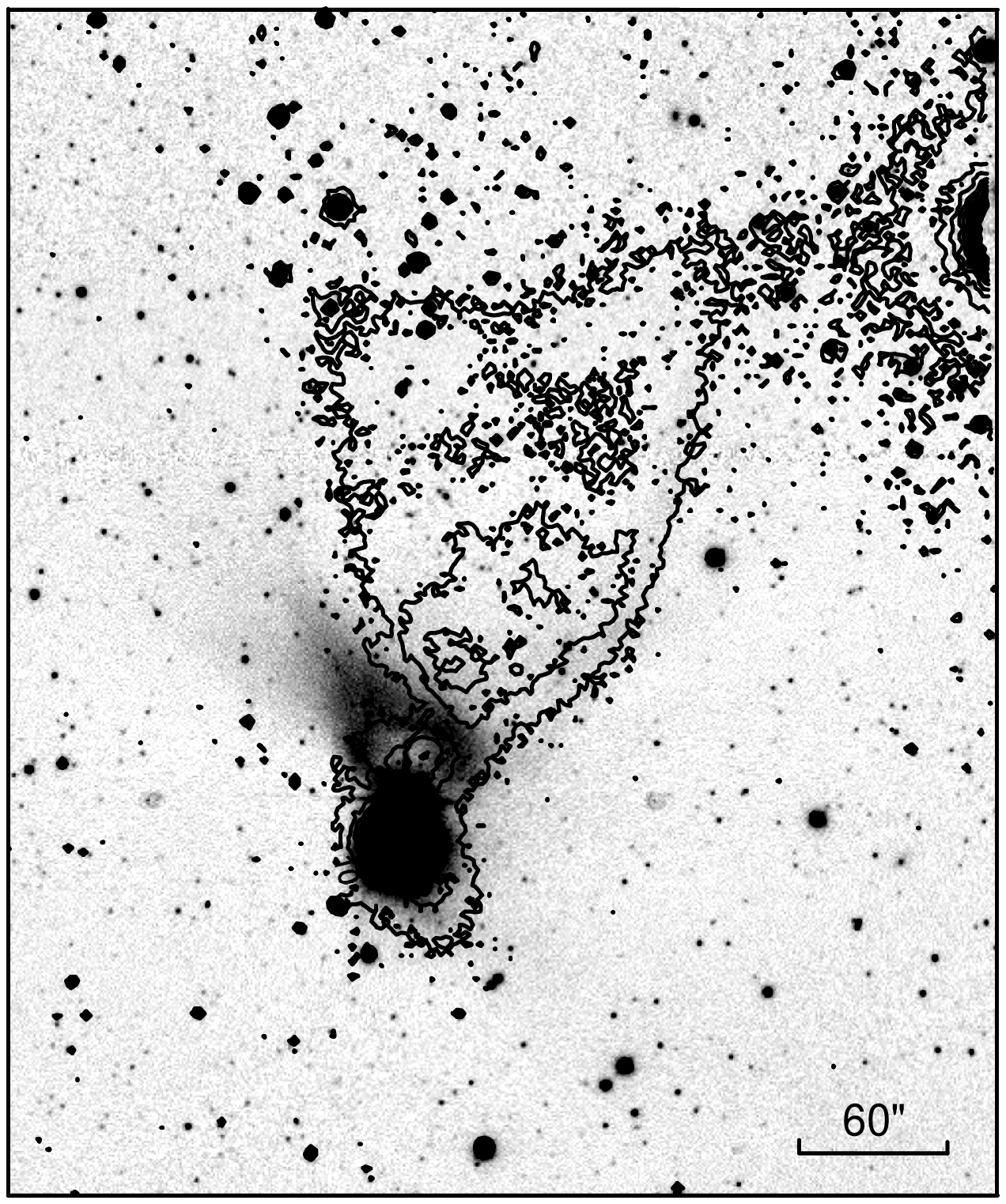}
\end{minipage}
\begin{minipage}{0.49\linewidth}
\includegraphics[width=0.74\linewidth]{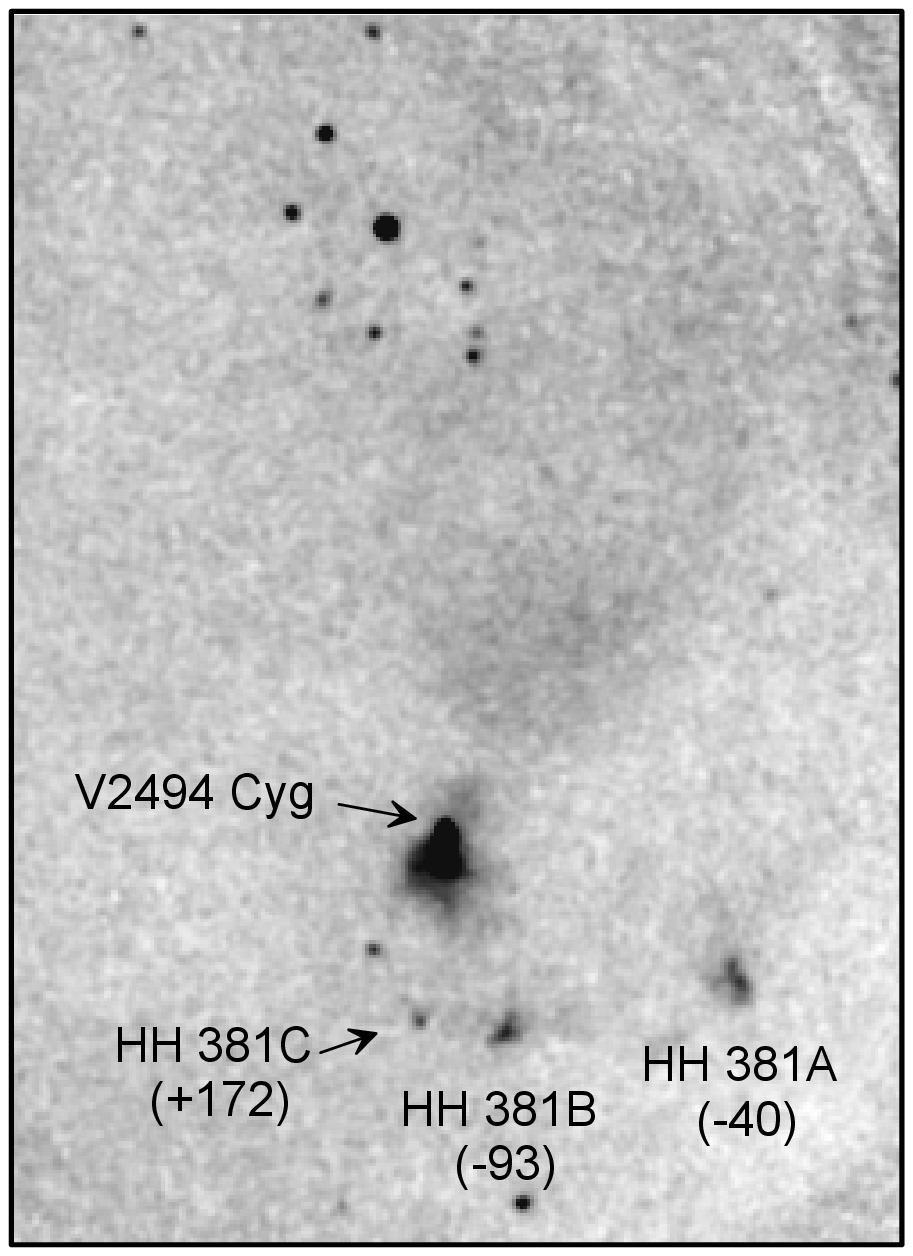}
\end{minipage}
\caption{Left panel: the superposition of the V2494~Cyg nebula images in K (gray scale) and in R$_{C}$ (isolines). Both images were obtained in
  2006 (see sec. 2.1 and 3.1 for the further details). Right panel: 
  the monochromatic H$\alpha$ image of the same field, restored from the FP data. The HH~381 knots A, B and C and their radial velocities in km sec$^{-1}$
are marked (see sec. 3.5). Both panels are in the same scale. }
\label{isolin}
\end{figure*}

The morphology of the nebula, associated with V2494~Cyg,  differs in the optical
and near-IR in much the same way as for the nearby Braid Nebula
\citep{movsessianetal1,movsessianetal2}. To make this comparison we
used the optical R$_{C}$ images obtained in 2006 with Suprime-Cam on
the Subaru telescope and the near-IR images obtained in the K-band
with WFCAM on UKIRT. These data are described in detail in the papers
of \cite{magakianetal2010} and \cite{khanzadyanetal}. For clarity  we present both images superposed in
Fig.\,\ref{isolin} (left panel).

To begin with, the orientation of the axis of the nebula is not easy
to define. At optical wavelengths the northern cone is more extended
but has a lower surface brightness and, as a whole, is aligned almost
north-south. However, its apex appears significantly detached and
shifted from the source star because of a dark lane that cuts across
it, producing between them another bright knot. The southern lobe
appears more inclined to the south-east; however, another narrow dark
lane near the central source with a position angle of 75\degr\ 
gives definitely greater deviation from the N-S direction. These
features  must be the result of foreground extinction variations, because, as
is clearly seen in Fig.\,\ref{isolin}a, in the near-IR the nebula is
symmetric and essentially oriented north-south. Note the total lack of
background stars in the optical range image to the east and west of the
reflection nebula (Fig.\,\ref{IRASneb}f), which shows that our object is located in the centre of the elongated dark cloudlet. Note also that in the optical range the western side of the northern cone has parabolic shape while its eastern side is nearly linear, being abruptly cut by the extinction. 

In the K-band image the northern cone is fainter, but its walls are
more clearly defined, especially the eastern side, which seems to be
fully absorbed in the optical image. The opening angle of the northern
lobe appears larger in the near-IR. Actually, the near-IR image seems brightest just in the places where the extinction in optical range is greatest. The southern lobe appears
significantly lower inclined to the east and has a much higher surface
brightness compared to the northern lobe. Thus, we can assume that the
real symmetry axis of the nebula is oriented much
closer to north-south than implied by the optical images.  This is in
full accordance with the direction of the parsec-size bipolar outflow
from the central star \citep{magakianetal2010}.

The spectacular biconic appearance of the nebula  points to its small inclination angle to the plane of sky. However, its morphology suggests that the northern cone should be inclined toward us; this conclusion is corroborated by the spectral data (see below).

A comparison of the DSS-2 and our recent Subaru images does not show
major morphological differences in the shape and brightness of the
nebula. 

\subsection{Photometry of the central star}

In order to attain a better understanding of the photometric history
of V2494~Cyg we collected brightness estimates from observations taken
at various epochs.

The first available image of the central star that we found was in the
POSS-1 survey red plate, obtained on 17 September 1952.  The source
magnitude in the USNO-B1.0 catalog \citep{monetetal} (where the
central star of the nebula is listed as object 1424-0432369) is given
as R1=18.57. Our measurements give nearly the same value. However, on
checking the image of V2494~Cyg in the SSS survey, which was performed
with better resolution than DSS-1, one can clearly see that even in
this epoch the object is extended and not purely star-like.  The
brightness of the central star may therefore be overestimated.
In any case the star definitely became brighter
in the 80-ies:
 from a Quick-V survey
plate (data obtained on 8 July, 1983) we measured the brightness to be
17.52 mag in V.

The blue plates from the Tautenburg collection, which span a period
from September 1975 to July 1985, show that the star is always below
the plate detection limit (which for various plates is estimated to be
in the range 17.20 to 20.58 mag in B).  

On DSS-2 plates the USNO-B1.0 magnitudes of the 1424-0432369 star are
as follows: B2=19.90; R2=13.74; IN=12.13. The errors on these values
are estimated to be about 0.25 magnitude. However, after further
analysis of the DSS-2 images and neighboring objects in the USNO-B1.0
catalogue, we came to conclusion that the measurements in R2 and IN
can be erroneous, because V2494~Cyg was recognised in the catalog as
an extended object, which led to a significant over-estimate of its
brightness. Indeed, on the DSS-2 R and I images all neighboring stars
with the same peak counts are cataloged in USNO-B1.0 with
significantly higher values of R2 and IN magnitudes, i.e. as much
fainter objects than V2494~Cyg.  Because of these factors we re-estimated the magnitudes of
the central star of the nebula on all images taken from the DSS and
SSS digital surveys, using the point-spread-functions (PSF) of field
stars. These results are included in the Table~\ref{RIphot}. 
 
To obtain recent photometric values as well as to look for
possible brightness variations we used several sets of images of
the object obtained since 2003. These include the 
IPHAS survey images \citep{drewetal}, our 2006 Subaru images, and the
observations from the 2.6-m telescopes in Byurakan and Crimea. All these
measurements also are included in Table~\ref{RIphot}. As can be
seen from Table~\ref{RIphot}, for more than twenty years the
brightness of the star has fluctuated by less than
0.3$^{m}$ amplitude from the mean value; its colour also remains nearly constant. There are
no signs of fading.

\begin{table}
\caption{The photometry of V2494~Cyg (HH~381~IRS)}
\label{RIphot}
\begin{tabular}{@{}llllll}
\hline 
Image, Epoch & B & V & R & I & R$-$I \\
\hline
   DSS-1 17.09.1952  & & & 18.7  &   \\
   Quick-V 08.07.1983 & & 17.5  \\
   DSS-2 24.08.1989  & 20.8  &   &  \\
   DSS-2 29.08.1989 & 20.3 & & \\
   DSS-2 24.08.1990 & & & 16.5  \\
   DSS-2 06.07.1991 & & & & 14.9 \\
   DSS-2 04.09.1991 & & & 16.3 \\
   DSS-2 03.07.1993 & & & & 14.6 \\
   IPHAS 14.11.2003 & & & 16.85 & 15.06 & 1.79  \\
   SUBARU 25/26.09.2006 & & & 16.52 & 14.91 & 1.61\\  
   ZTSh 22.05.2008  & & & 16.29  & 14.51  & 1.78 \\
   ZTSh 16.06.2008 & & & 16.36  & 14.56  & 1.80 \\
   ZTA 25.06.2008  & & & 16.29  & 14.56  & 1.73  \\
   ZTA 06.10.2008  & & & & 14.49 & \\
   ZTA 24.06.2009  & & & 16.32  & 14.71  & 1.61  \\
   ZTA 28.06.2009  & & & 16.09  & 14.47  & 1.62  \\
   ZTA 06.11.2009  & & & 16.15  & 14.49  & 1.66  \\
   ZTA 06.08.2010  & & & 16.34  & 14.76  & 1.58  \\
   ZTA 02.11.2010  & & & 16.39  & 14.57  & 1.82 \\
\hline
\end{tabular}
\medskip

Photometric data for DSS-1 are taken from USNO-B1.0 catalog; all other values are our own estimates (see text).
\end{table}

In summary, we conclude that the increase in brightness of the V2494~Cyg very probably started in the early 1980s; in the
period between 1990 and 2000 the star already reached its maximum
visible brightness and since then has remained at this luminosity. The absence of  the star on the Tautenburg plates (even though in B it should be very faint) gives some support to this conclusion in the sense that there were no major outbursts before 1983. The
amplitude of 2.5 mag in R is probably a lower limit, since even at the
time of the POSS-1 survey observations the star was slightly nebulous
and therefore extended. In fact, on the base of our preliminary
reports the object is already recognized as a variable star and numbered as
V2494~Cyg \citep*{kazarovetsetal}.  We suggest that in future only this
name be used for the central source within the nebula.  Note in
particular that the star actually is not associated with HH~381 (see below), so
the name HH~381-IRS should certainly be avoided.  The nebula in turn
can be referred to as ``the V2494~Cyg nebula''; on the other hand, ``the IRAS~20568+5217 nebula''
name also could be kept for the identification purposes (it was
named as such when first detected by \citealt{devineetal}.).
\begin{table}
\caption{The log of spectral observations}
\label{specobs}
\begin{tabular}{@{}lll}
\hline
Date & Object & Telescope, system \\
\hline
   10.01.2007 & V2494 Cyg (star+nebula) & 6-m, SAO, long-slit \\
   13.06.2008 & HH 382A  & 2.6-m, Byurakan, long-slit \\
   29.06.2008 & V2494 Cyg field & 6-m, SAO, scanning FP \\
\hline
\newline\end{tabular}
\end{table}

\subsection{Slit spectroscopy of the central star, nebula and bipolar jet}

In the slit spectrogram, described above, the spectra of the central
star and of the neighboring parts of nebula were registered.

\begin{figure}
\centerline{\includegraphics[width=15pc,angle=270]{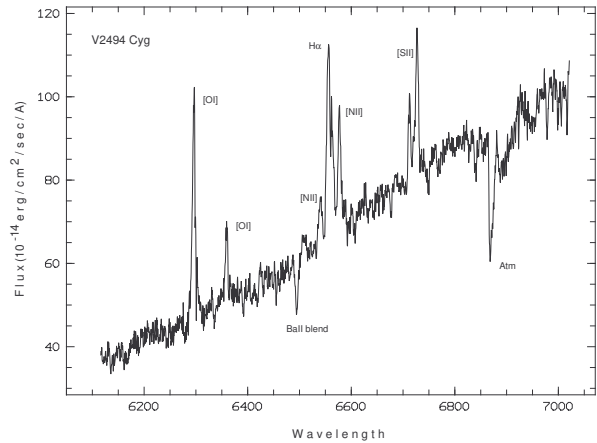}}
\caption{The optical spectrum of the V2494~Cyg in red range, obtained
  with 6-m telescope. All major features are identified.}
\label{star_spec}
\end{figure}

The spectrum of V2494~Cyg, extracted by using the stellar
PSF, is shown in Fig.\,\ref{star_spec}. We see the reddened stellar continuum with prominent emission lines
(H$\alpha$, [OI], [NII] and [SII]) superimposed. Such strong emission
lines, unusual for FUors, almost certainly belong to the collimated,
ionized jet that is visible in published direct images
\citep{magakianetal2010}.  

The only strong stellar absorption, which is evident in Fig.\,\ref{star_spec}, is  the strong Ba~II blend near
$\lambda$6495. This feature is typical of G supergiants and thus distinctive
for FU~Ori type spectra \citep{reipurthetal2002}. As can be seen from the Fig.2 and Fig.4 of that paper, there are no other conspicuous absorption
features, which one might expect to see in the observed wavelength
range with our spectral resolution. As for the  P~Cygni absorption component in the H$\alpha$
line,   which are usually prominent in the spectra of
FU~Ori type stars,
it is very probably masked by the rather strong emission line of the jet (see, however, sec 3.4).  

\begin{figure*}
\centerline{\includegraphics[width=26pc]{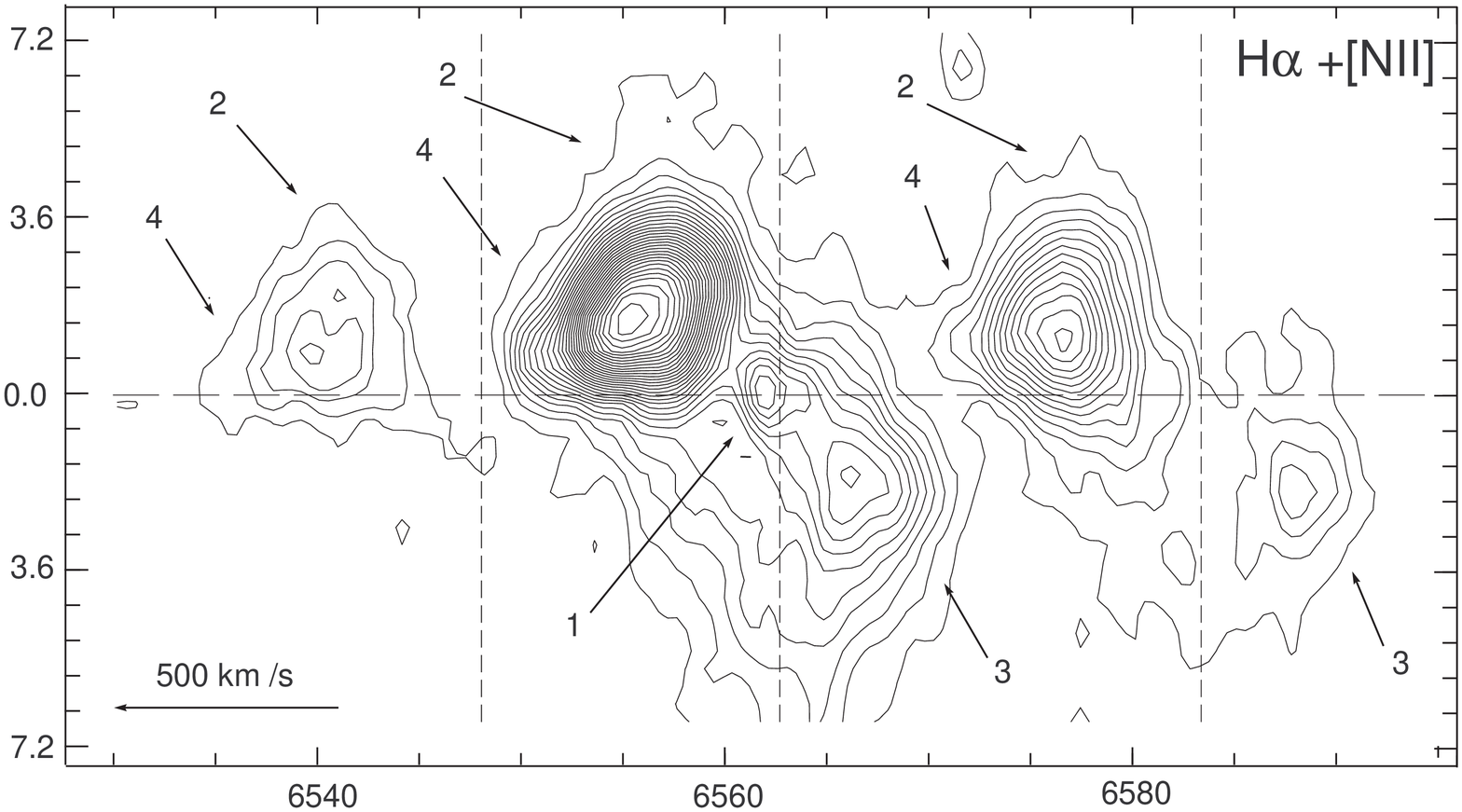}}
\centerline{\includegraphics[width=22pc]{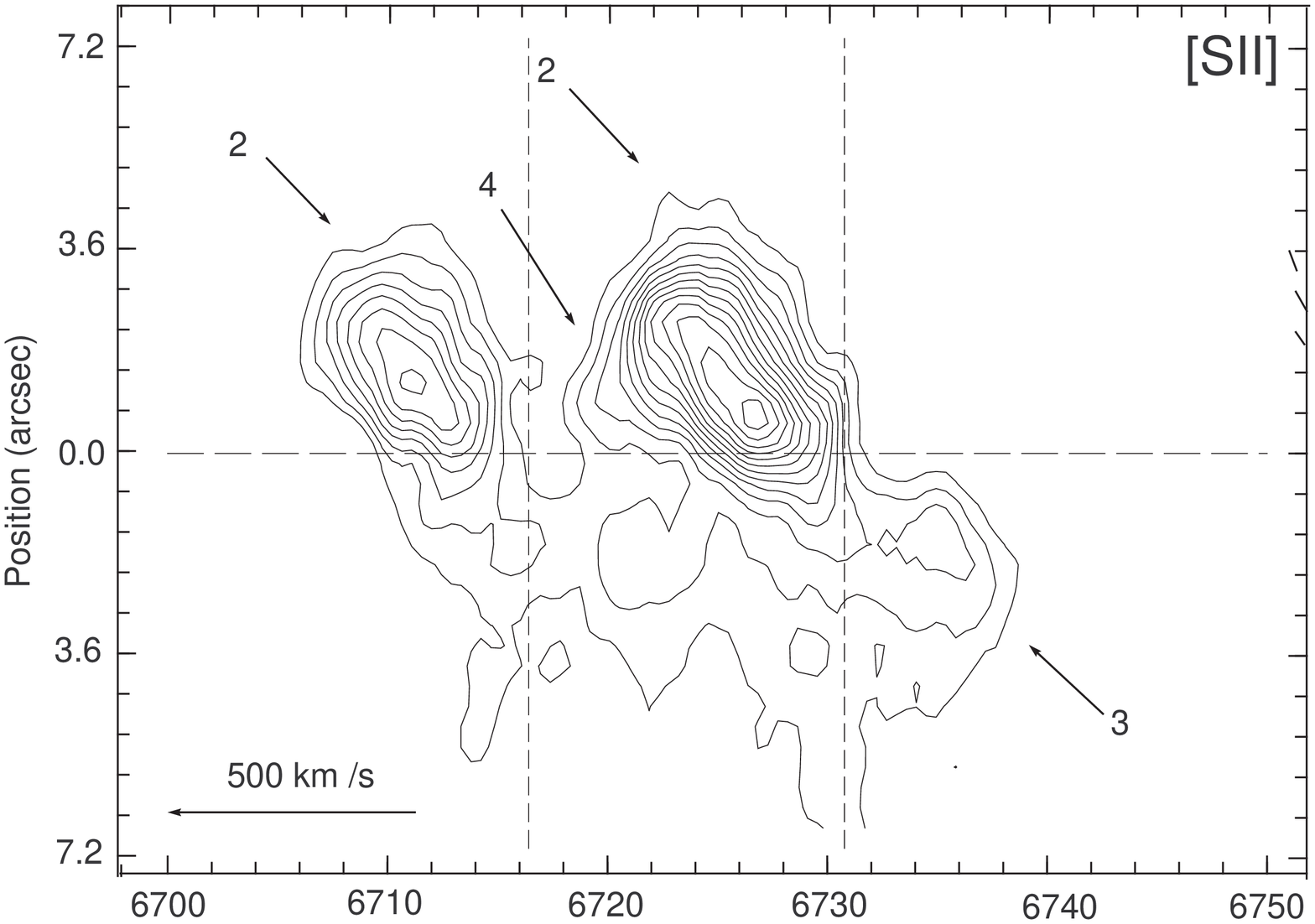}}
\centerline{\includegraphics[width=35pc,height=17pc]{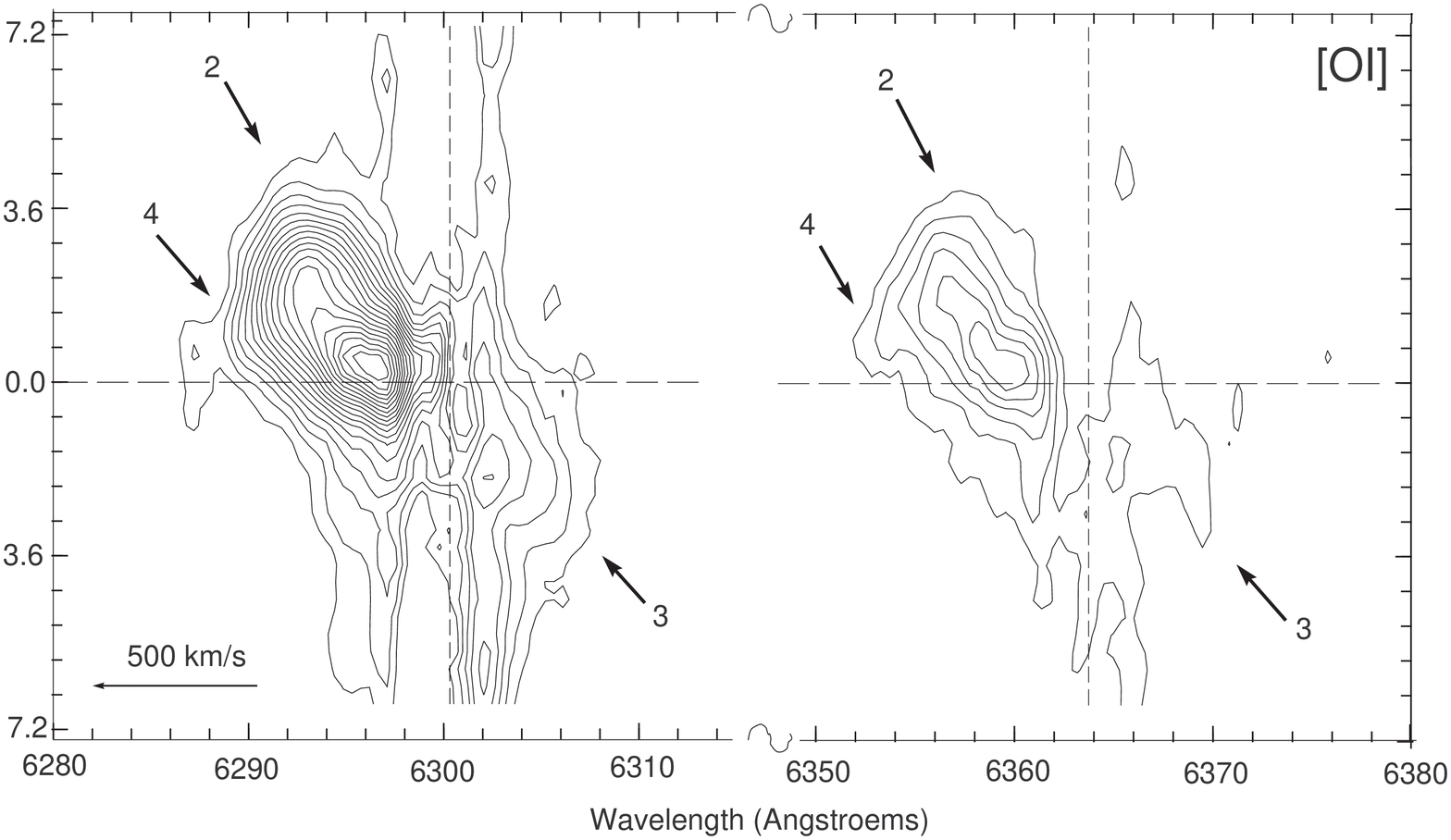}}
\caption{Continuum-subtracted long-slit spectrum of the V2494~Cyg star
  and its vicinities, showing the emission lines of H$\alpha$ and
  [NII] (upper panel), [SII] (middle panel) and [OI] (lower panel;
  vertical artifacts are produced in the process of the bright night
  sky lines subtraction). Horizontal dashed lines in each panel
  correspond to the stellar position, and the vertical ones mark the
  laboratory wavelengths. The slit orientation is
indicated in Fig.\,\ref{refl_Ha} (right panel). Several discernible components in H$\alpha$
  and other emission lines are denoted by numbers: 1 -- probable
  stellar emission; 2 -- the jet emission; 3 -- the counter jet
  emission; 4 -- a component with very high negative velocity (see
  text for their discussion and radial velocities).}
\label{JetSpecs}
\end{figure*}

The observed emission lines, especially H$\alpha$, have complex
profiles and are extended along the slit. To investigate the emission
spectrum in more detail we have subtracted the continuum from our
reduced spectral image, separately fitting and subtracting the stellar
spectrum row by row. The result is shown in Fig.\,\ref{JetSpecs}. The slit orientation is
shown in Fig.\,\ref{refl_Ha}. In
Fig.\,\ref{JetSpecs} we see that the intensity maxima of all main emission lines are
shifted to the north with respect to the stellar continuum. This fully
confirms that the emission originates in the ionized jet that extends
to the north from the central star \citep{magakianetal2010}. In
Fig.\,\ref{JetSpecs} the jet emission features are labelled with the
number 2; signs of the counter jet extending to the south are also
evident in the data, and are labelled 3 in Fig.\,\ref{JetSpecs}.

As is also evident in Fig.\,\ref{JetSpecs}, all emission lines
demonstrate complicated spectral and spatial structure. The general
trend of a rapid increase in radial velocity (all radial velocities in this paper are heliocentric) with distance from the
source (a possible indication of acceleration) is clearly seen in the
[SII] and [OI] profiles. The peak velocities in these lines increase
from $-$170 and $-$197 km s$^{-1}$ near the star to $-$305 and $-$348
km s$^{-1}$ at a distance of 2.5 arcsec. In comparison, the [NII]
profiles in Fig.\,\ref{JetSpecs} are not inclined, though they do
exhibit high negative velocities: the [NII] radial velocity measures
$-$313 km s$^{-1}$ at the star position and remains at values of about
$-$320 km s$^{-1}$ along the observed length of the jet. We also note
that all of the forbidden line profiles are wide, with Full Width Half
Maximum (FWHM) widths of about 250 km s$^{-1}$ (in [SII] FWHM
$\sim$ 340 km s$^{-1}$).  The profiles also consist of at least two
components separated by about 80 - 100 km s$^{-1}$. However, our
spectral resolution is not sufficient for a reliable multi-component
fitting of these profiles.

As was mentioned above, the structure and behavior of the H$\alpha$
emission profile is the most complicated of all lines observed. The
same spatially extended jet component with a near-constant velocity
(peak velocity measures $-$320 km s$^{-1}$ at the star position and
$-$290 km s$^{-1}$ near the observed end of the jet; a similar
velocity is observed here in [NII]) is evident in
Fig.\,\ref{JetSpecs}, in addition to a high-velocity component very
close to (but not coinciding with) the driving source. This component
is labelled 4 in Fig.\,\ref{JetSpecs} and can be traced from 0.4
to 1.1 arcsec along the northern flow lobe.  We estimate its mean
velocity to be $-$510 km s$^{-1}$, while its blue wing reaches $-$645
km s$^{-1}$.  Careful inspection of our data shows that this component
can also be seen in forbidden lines, with a mean velocity of about
$-$430 km s$^{-1}$ (in good agreement with the H$\alpha$ results).
The existence of this high-velocity component in $\lambda$6730
emission combined with its absence in $\lambda$6717 [SII] emission
indicates a high density for the emitting medium. We are inclined to
consider this feature as a separate component unrelated to the very
broad, single-peaked H$\alpha$ because it is not symmetric -- neither
spectrally (it has the appearance of a rather extended emission wing
from the blue side of a strong and more sharp core) nor spatially (the
brightest H$\alpha$ emission is offset by about 2 arcsec to the north
from the broadest part of the profile).

Finally, we label with number 1 in Fig.\,\ref{JetSpecs} a discrete
low-velocity peak seen in H$\alpha$ that is centred at about $-$20 km
s$^{-1}$. This component is spatially coincident with the maximum of
the stellar continuum and is not extended along the slit length (i.e.
in a north-south direction).  The FWHM of this feature is about 140 -
160 km s$^{-1}$.  With respect to the local standard of rest (LSR), this
component is blue-shifted by less than $-$12 km s$^{-1}$. To compare, the systemic LSR velocity of the whole cloud is $-$3.4 km s$^{-1}$ (Moriarty-Schieven et al., in preparation). We therefore identify this component with the star itself. Weak,
low-velocity H$\alpha$ emission lines are typical of FU~Ori objects
(e.g. \citealt{herbigetal}). These are probably too narrow to be
associated with accretion (see for example the much wider permitted
HI lines observed towards T~Tau stars: \citealt{folhaemerson});
instead, this emission feature is probably of chromospheric origin.
Moreover, the absence of this feature in the observed forbidden line
profiles corroborates the conclusion that all strong forbidden
emission lines (which are typically absent or very weak in the spectra
of classical FUors) belong only to an associated outflow.

The counter jet associated with V2494~Cyg is much fainter in our data,
though is nonetheless detected. Emission from the counter jet can be
traced 4.3 arcsec to the south of the stellar source continuum.  Because
of the faintness and complexity of the emission (especially in [SII]),
radial velocities are not so easy to measure.  Even so, by averaging
all lines, we estimate a typical counter jet velocity of $+$163 $\pm$
39 km s$^{-1}$.  A velocity gradient similar to that seen in the
northern lobe cannot be unequivocally identified,  although hints of it can be seen in the [SII] and [OI] profiles.

The electron density in the northern jet lobe, estimated from the
[SII] line ratio, is about 7000 cm$^{-3}$; in the southern counter jet
a density of around 5000 cm$^{-3}$ is measured.  Within the high
velocity component densities of more than 10000-11000 cm$^{-3}$
are estimated.

\subsection{ Scattered light spectroscopy}

\begin{figure*}
\centerline{\includegraphics[width=38pc,angle=0]{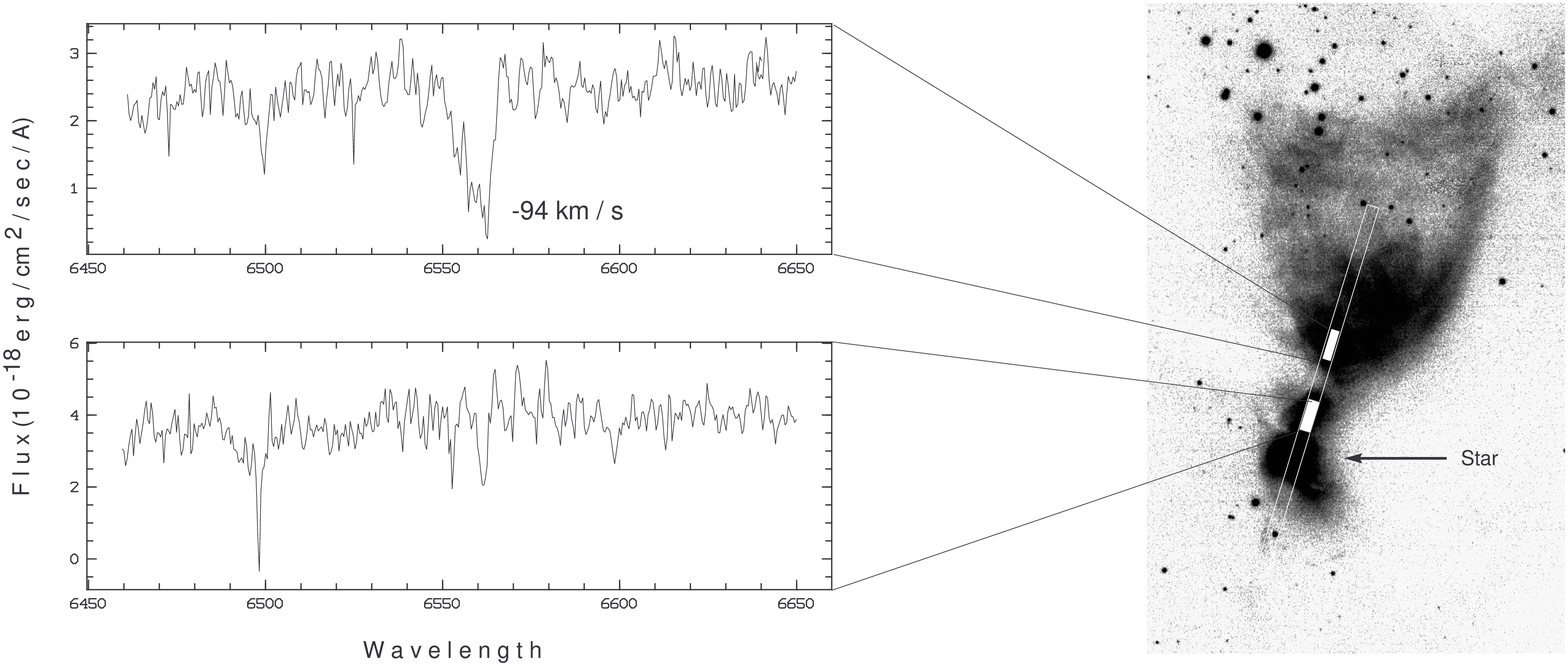}}
\caption{The H$\alpha$ line profile as seen in the reflection in the
  distance about 10-15 arcsec (lower panel) and 35-40 arcsec (upper panel) from the star. To increase the S/N ratio
  several rows of the nebular spectrum are averaged. Note that the
  blue wing of the asymmetric absorption in the upper panel reaches
  about $-$700 km s$^{-1}$ in radial velocity. }
\label{refl_Ha}
\end{figure*}

As was noted earlier, we were not able to detect any signs of
H$\alpha$ absorption in the stellar spectrum, which is probably
masked by very strong and broad jet emission.  However, during our analysis of the
reflected spectrum of the nebula, we quite unexpectedly found that at
a distance of about 25 - 50 arcsec to the north of the star
the nebula spectrum possesses a strong, broad, and blueshifted
H$\alpha$ absorption feature (Fig.\,\ref{refl_Ha}). This feature is
centred at a velocity of $-$94 km s$^{-1}$ and exhibits a blue wing
that extends up to $-$700 km s$^{-1}$. Such a profile is typical of
FU~Ori objects.  As is shown in Fig.\,\ref{refl_Ha}, this absorption
feature is observed toward the northern lobe of the conical reflection
nebula while, closer to the source, it is virtually absent.  In
Fig.\,\ref{refl_Ha} we show, for example, a spectrum extracted from
the bright knot located at a distance of 6.5 - 15 arcsec from the
star (this knot is closer to the central star though is nevertheless
far beyond the emission jet).

With this discovery V2494~Cyg and its nebula become one more example
of a source exhibiting the rare phenomenon known as ``spectral
asymmetry''. This was first reported in observations of the R~Mon
$+$ NGC~2261 system
\citep*{greenstein1948a,greenstein1948b,stocktonetal,greensteinetal}.
This asymmetry was successfully explained only in the work of
\citet{jonesherbig}.  We will discuss this unusual feature in more
detail below.  In the meantime, the detection of the wide P~Cyg type
absorption feature again confirms the FU~Ori nature of V2494~Cyg.

\subsection{ Spectroscopy of HH knots}

As was described above, V2494~Cyg is also referred to as HH~381~IRS
because it was believed to be the driving source of nearby HH~381 group of HH objects \citep*{devineetal}. However,
the relationship of the knots that comprise HH~381 to each other and to
V2494~Cyg itself was not obvious until the our recent observations. 

In the Fig.\,\ref{isolin} (right panel) we show the monochromatic H$\alpha$ image of the field around  V2494~Cyg, which is reconstructed from the our FP data. This should be compared with  the Fig.4 in \cite{magakianetal2010} paper.  Our data confirm
the emission nature of the HH~381 knots A, B and C
and do not show any other nebulous emission
objects close to V2494~Cyg. Observed H$\alpha$ emission line profiles
are single-peaked in all three knots, although they differ in width:
the FWHM is about 50 km\,s$^{-1}$ for
knots A and B and about 100 km\,s$^{-1}$ for knot C. The radial
velocities of these knots differ even more drastically. While knots A
and B have V$_r$ = $-$40 km\,s$^{-1}$ and $-$93 km\,s$^{-1}$
respectively, the radial velocity of HH~381 C is +172
km\,s$^{-1}$ (see also  right panel of the Fig.\,\ref{isolin} to compare the relative positions of the HH knots). This last  value ideally corresponds to the velocity of the
V2494~Cyg counter jet (see sec. 3.3) and actually confirms that HH~381~C
indeed belongs to the southern lobe of the V2494~Cyg outflow, along with the fainter HH knots HH~382 E, F, G \citep{magakianetal2010} and the
disrupted HH~382 A-D group at the head of the southern outflow. All of
these objects lie almost precisely on a straight line that coincides
with the axis of the nebula. However, knots
HH~381 A and B must belong to some other outflow, as has previously
been suspected based on their location. This is further supported by
the morphology of the knots: HH~381~C looks like a small bow shock
heading southward, while the shapes of HH~381 A and B do not show
marked symmetry.
 
As was mentioned above, we were able to obtain the slit spectrum of knot
HH~382~A. The spectrum is typical of HH objects, with H$\alpha$, [SII]
and [OI] emission and no traces of continuum. The line intensities
correspond to low density and moderate excitation. The radial velocity
of this knot is rather low (+23 $\pm$ 17 km\,s$^{-1}$) but is positive,
which confirms that the HH 382~A-D group is the leading part
of the southern flow (at least in the optical range). Taking into
account its distance from the source and disrupted appearance, such
a low velocity is not surprising.

\section{Discussion}
\subsection{The Star}

The central star V2494~Cyg was detected in the 2MASS survey as the
object 2MASS J20582109+5229277 with the following photometric values
in the three near-infrared bands: J=11.54$\pm$0.031, H=9.81$\pm $0.030
and K=8.31$\pm$0.017 mag. The estimate of the interstellar extinction
$(A_{V} = $ 5.8 mag), taken from the map of \cite{rowlesfroebrich},
places V2494~Cyg in the locus of T~Tau stars in the two-colour
diagram \citep{rydgrenvrba,meyercalvet}. Even without the application
of the extinction correction we see that V2494~Cyg is located among
other FU~Ori-like objects in this diagram (see
\citealt*{greeneaspinreipurth} and \citealt{aspinetal}). But the bolometric
luminosity of V2494~Cyg, estimated from IRAS fluxes
\citep{reipurthetal1993}, is quite low for a FUor. The distance
controversy leads to a wide range of estimates: from 14
L$\sun$\ (D=700 pc, \citealt{reipurthaspin1}) to 45.55 L$\sun$\ (D=1300
pc, \citealt*{connelleyetal}). For 800 pc, as is assumed in our works,
this approach yields 18 L$\sun$.

Nevertheless, there are many arguments in favour of V2494~Cyg,
formerly HH~381~IRS, being a genuine FU~Ori type object.
\begin{description}
\item 
Its infrared spectrum is virtually identical to other FUors \citep{aspinetal}.
\item
Its optical spectrum shows wide and blueshifted H$\alpha$ absorption
(though seen only in reflection as described in sec. 3.4) and some
other features typical of FUors.
\item
Its increase in optical brightness was actually observed; it remains
at maximal brightness for more than 20 years.
\item
 Its spectral type can be estimated as G
(in the region near H$\alpha).$ 
\item
There is no detected Fe~II or Fe~I emission in the optical spectrum,
typical for T~Tau stars and EXors.
\end{description}

Nevertheless, several properties of this star make it
a rather unusual FUor.
\begin{description}
\item
The amplitude of the outburst is only $\approx2.6$ mag in R (however,
if the R1 magnitude in USNO-B is also overestimated, it will make the
amplitude larger and more similar to classical FUors).
\item
Its post-outburst bolometric luminosity is quite low.
\item
The object is the source of a giant parsec-sized bipolar flow.  On the
other hand, even if some of the known FUors are related to HH objects,
only a few of them possess jets and develop extended outflows.
\item
Rather strong forbidden emission lines are observed in the spectrum of
the star. Even if we assume that they all originate only in the jet,
this is not typical of FUors.
\end{description}

In fact, V2494~Cyg is already included in the list of 10 known FU Ori
type objects with the outburst detected \citep[ V2494~Cyg is referred
  to as HH~381~IRS in this paper]{reipurthaspin3} but among them there
are no other objects with so strong emission. We will return to this
question below.

Another nearly unique feature of V2494 Cyg and its nebula is the pronounced ``spectral asymmetry'' (by which we mean that the spectra of the star observed directly and through
reflection show marked differences). This asymmetry is the direct result of the
existence of the anisotropic expanding stellar envelope around the
central star combined with the geometrically favorable projection of
its spectrum on the walls of the cavity in the interstellar dust (we
see these walls as the bright conical nebula, which reflect the
spectrum of the envelope at various latitudes in the line-of-sight
direction). Besides the classic example of R~Mon \citep{jonesherbig},
more or less well studied cases include RNO~129
\citep{movsessianmagakian} and PV~Cep \citep{movsessianetal3}. The
object V2494~Cyg, however, is the first FUor in this group, which
makes it even more intriguing. First of all, it confirms that we
observe this star near its equatorial plane where the influence of the
outflowing envelope is minimal; this implies that
FUors with the most prominent P~Cyg absorptions should
be oriented nearly pole-on. In addition, it gives a chance to
analyze the structure of the envelope, similarly to R~Mon
\citep{magakianmovsessian}, and to compare it with theoretical
predictions. Spectropolarimetric studies of this nebula will be very
useful in separating the reflected and \textit{in situ }produced
spectral lines.

\subsection{Outflow}

The outflow of V2494~Cyg can be considered to be a typical
parsec-sized outflow. However, such long flows are unusual for
FUors. Among all known FUors and FUor-like objects only L~1551~IRS5
(which most closely resembles V2494~Cyg, also taking into account the
very low luminosity of both central stars), Z~CMa (although it is not
known yet if this flow is produced by the FUor or its more massive
companion) and the recently found neighbor Braid star (also called
HH~629~IRS or V2495~Cyg) possess such well-developed optical
flows. Furthermore, this outflow is the most extended of all those
listed above - more than 7 pc in full span in the optical. With its
probable molecular hydrogen extensions, described in detail by
\cite{khanzadyanetal}, it may extend even to 9 pc. Here we should
mention also that V2494~Cyg drives an extended N-S molecular
outflow, detected in the radio wavelength range (Moriarty-Schieven et
al. 2013, in preparation).

It is obvious that such a long flow should have a significant age,
which can also be inferred from the disrupted morphology of its
leading bow shocks \citep{magakianetal2010}. In this previous work,
the kinematic age of 17,600 years was given for its most distant
optical bow shock HH~967. This estimate was based on the assumption of
200 km\,s$^{-1}$ for the mean velocity of the outflow and 800 pc for
the distance. For the preceding molecular hydrogen clumps the age will
exceed 20,000 years. However, such age estimates for the V2494~Cyg
flow are very approximate because, on the one hand, the ejection
velocity can be far underestimated (see Sect. 3.3) and, on the other
hand, the outflow probably decelerates (Sect. 3.5).  In any case, the
age of the V2494~Cyg outflow should be at least about 10,000 years
(probably much more: in the paper of \citealt{khanzadyanetal} it is shown
that it can reach even 10$^{5}$ years).  Thus, it originated long
before the detected recent outburst.  Just the same situation exists
for the neighbor V2495~Cyg (formerly known as the Braid star) and its
outflow \citep{khanzadyanetal}.

One should keep in mind that V2494 and V2495~Cyg are so far the only
known FUors with observed recent eruptions \textit{and} extended
high-age flows.  However, these remarkable examples still cannot differentiate between various mechanisms which have been suggested to
explain the FU~Ori phenomenon. The hypothesis that FU~Ori outbursts
should be directly connected with extended outflow formation, proposed
several tens of years ago \citep{dopita,reipurth} seems very
attractive \citep[see also][]{reipurthaspin3}. Yet we actually have no
direct evidence: extended outflows can be the independent
manifestations of the strong accretion during PMS stellar evolution.
However, in any case multiple paired knots can be considered as
evidence of the periodic eruptive character of mass-loss events
\citep{reipurthbally}. The recent analysis of the morphology of nearly
30 molecular outflows by \cite{ioannidisfroebrich} confirmed the
estimates, made from optical studies, namely that the typical time
gaps between the significant ejections in the outflow are of the order
of 10$^{3}$ years. Meanwhile, the time between recurrent FU~Ori
outburst events is believed to be about 10$^{4}$ years
\citep{hartmannkenyon,reipurthbally}. Hence, they seem in closer
agreement with the total outflow lifetime.  However, the absence of an
object with two observed recurrent FU~Ori outbursts makes all these
indirect estimates highly speculative.

Nevertheless, the discovery of a jet in the immediate vicinity of
V2494~Cyg makes the situation even more intriguing.  Estimating from
our spectrograms the visible length of the jet (and counterjet) as
4 arcsec, we get its kinematical age as about 40 years (for the 400
km\,s$^{-1}$ as ejection velocity). This value is in the good agreement with our
estimation of the probable date of the eruption (with all the caution about the uncertainties both in the eruption date and the jet age) and, in turn, implies
that the jet started to propagate near the moment of the V2494~Cyg
outburst. To our knowledge, such a direct connection is found here for
the first time. It can be considered as the first direct proof that
the FU Ori outbursts create collimated jets (at least in some cases: of course, we cannot state that\textit{ all} FU Ori events produce directed outflows).  Taking into account the
rather large kinematical age of the full V2494 Cyg outflow, one can go
further and suggest that its distant and disrupted bows were created
during one or even several (depending on the age estimates) previous
FU~Ori events in V2494~Cyg, thus making this star the first proven
recurrent FU~Ori object.

Another interesting feature is the different velocity gradients for
the lines of higher and lower excitation level in the jet and the
counterjet. The more or less similar behavior of emission lines was
found in the small HH knot HH~214 in the GM~1-27 nebula
\citep{magakianmovsessian1995}.  One can suggest that the H$\alpha$
and [NII] emission traces the high-velocity/high-density axis of the
jet, while the [SII] and [OI] lines are excited in entrained gas that
is steadily accelerated by the jet.  Of course, other explanations are
not excluded. Yet another feature of interest is the definite
multi-component structure of the forbidden line profiles.  This
perhaps can be understood if we assume that the jet itself consists of
several tiny knots of various excitation level which are not yet well
separated.  The existence of the very high velocity and high density
component just within 1 arcsec of the star supports this suggestion
to some extent. It can even be the indication of a new ejection. A
similar single high-velocity knot, seen in forbidden lines, was
detected in the outflow of LkH$\alpha$~324SE \citep{herbigdahm}.

Concerning the origin of the HH~381 A and B knots, their radial
velocity data definitely reject their affiliation with the extended
N-S flow from V2494~Cyg. Thus, they likely belong to some other
flow. One still can return to the suggestion made in our previous
paper \citep{magakianetal2010}, that these knots are either produced
by the hypothetical companion of V2494~Cyg (the only argument for this
is their proximity) or that they correspond to the western lobe of the
extended flow from IRAS~20573+5221.

We reach the conclusion that the combination of all the
features discussed above indeed makes V2494 Cyg a unique object even
among so rare a class as FUors.

\section*{Acknowledgments}

This work is based on observations collected at Subaru
  observatory, UKIRT, Special Astrophysical Observatory (Russia) and
  Byurakan observatory (Armenia). 

We thank Alexandr Burenkov for the observations with the 6-m telescope, Helmut Meusinger for the analysis of the plates from the
Tautenburg collection and Konstantin Grankin for obtaining the images
from the Crimean Observatory.

The observations with the 6-m telescope of the Special Astrophysical
Observatory of the Russian Academy of Sciences (SAO RAS) were carried
out with the financial support of the Ministry of Education and
Science of Russian Federation (contracts no.~16.518.11.7073 and
14.518.11.7070).

We are grateful to the referee, whose many suggestions and corrections helped to improve the paper.

The authors wish to recognise and acknowledge the very significant
cultural role and reverence that the summit of Mauna Kea has always
had within the indigenous Hawaiian community. We are most fortunate to
have the opportunity to conduct observations from this sacred
mountain.

The Two Micron All Sky Survey is a joint project of the University of
Massachusetts and the Infrared Processing and Analysis Center, funded
by NASA and the NSF.  The Digitized Sky Surveys were produced at the
Space Telescope Science Institute under U.S. Government grant NAG
W-2166.  This research has made use of data obtained from the
SuperCOSMOS Science Archive, prepared and hosted by the Wide Field
Astronomy Unit, Institute for Astronomy, University of Edinburgh,
which is funded by the UK Science and Technology Facilities Council.

This publication makes use of the extinction map query page hosted by
the Centre for Astrophysics and Planetary Science at the University of
Kent.

This work was made possible in part by a research grant astroex-3124 from the Armenian National Science and Education Fund (ANSEF) based in New York, USA.

\label{lastpage}


\begin{thebibliography}{99}
\bibitem[\protect\citeauthoryear{Afanasiev \& Moiseev}{2005}]{afanasievmoiseev} Afanasiev V. L., Moiseev A. V., 2005, Astron. Let, 31, 194
\bibitem[\protect\citeauthoryear{Aspin et al.}{2009}]{aspinetal} Aspin C., Beck T. L., Pyo T.-S., Davis C. J., Schieven G. M., Khanzadyan T., Magakian T. Yu., Movsessian T. A., Nikogossian E. G., Mitchison S., Smith M. D., 2009, AJ, 137, 431
\bibitem[\protect\citeauthoryear{Aspin et al.}{2011}]{aspinetal2011} Aspin C., Beck T. L., Davis C. J., Froebrich D., Khanzadyan T., Magakian T. Yu., Moriarty-Schieven G. H., Movsessian T. A., Mitchison S., Nikogossian E. G., Pyo T.-S., Smith M.D., 2011, AJ, 141, 139
\bibitem[\protect\citeauthoryear{Bell \& Lin}{1994}]{belllin} Bell K. R.,  Lin D. N. C., 1994, ApJ, 427, 987
\bibitem[\protect\citeauthoryear{Bonnell \& Bastien}{1992}]{bonnellbastien} Bonnell I., Bastien P., 1992, ApJ, 401, L31
\bibitem[\protect\citeauthoryear{Connelley, Reipurth \& Tokunaga}{Connelley et al.}{2007}]{connelleyetal} Connelley M. S., Reipurth B., Tokunaga A. T., 2007, AJ 133, 1528
\bibitem[\protect\citeauthoryear{Devine, Reipurth \& Bally}{Devine et al.}{1997}]{devineetal} Devine D., Reipurth B., Bally J., 1997, in Malbet F., Castets A., eds, Poster proc. IAU Symp. 182, Low Mass Stars Formation - from Infall to Outflow, Observatoire de Grenoble, p. 91
\bibitem[\protect\citeauthoryear{Dobashi, Bernard \& Fukui}{Dobashi et al.}{1996}]{dobashietal1} Dobashi K., Bernard J.-P., Fukui Y. 1996, ApJ, 466, 282
\bibitem[\protect\citeauthoryear{Dopita}{1978}]{dopita} Dopita M. A., 1978, ApJS, 37, 117
\bibitem[\protect\citeauthoryear{Drew et al.}{2005}]{drewetal} Drew J. E., Greimel R., Irwin M. J. et al., 2005, MNRAS, 362, 753
\bibitem[\protect\citeauthoryear{Folha \& Emerson}{2000}]{folhaemerson} Folha D. F. M., Emerson, J. P., 2000, A\&A, 365, 90
\bibitem[\protect\citeauthoryear{Garrido et al.}{2002}]{garridoetal} Garrido O., Marcelin M., Amram P., Boulesteix J., 2002, A\&A, 387, 821 
\bibitem[\protect\citeauthoryear{Greene, Aspin \& Reipurth}{Greene et al.}{2008}]{greeneaspinreipurth} Greene T. P., Aspin C., Reipurth B., 2008, AJ, 135, 1421
\bibitem[\protect\citeauthoryear{Greenstein}{1948a}]{greenstein1948a} Greenstein J. L., 1948a, Harvard obs. Monographs No. 7 (Proc. of Cennential Symp.), p. 19
\bibitem[\protect\citeauthoryear{Greenstein}{1948b}]{greenstein1948b} Greenstein J. L., 1948b, ApJ, 107, 375
\bibitem[\protect\citeauthoryear{Greenstein et al.}{1976}]{greensteinetal} Greenstein, J. L., Kazarian M. A., Magakian T. Yu., Khachikian E. Ye., 1976, Astrophysics, 12, 384
\bibitem[\protect\citeauthoryear{Hartmann}{2009}]{hartmann2009} Hartmann L., 2009, Accretion Processes in Star Formation, Cambridge Univ. Press (2nd edit.), chapt. 9, p. 188
\bibitem[\protect\citeauthoryear{Hartmann \& Kenyon}{1996}]{hartmannkenyon} Hartmann L., Kenyon S. J., 1996, ARA\&A, 34, 207
\bibitem[\protect\citeauthoryear{Herbig}{1977}]{herbig2} Herbig G. H., 1977, ApJ, 217, 693
\bibitem[\protect\citeauthoryear{Herbig \& Dahm}{2006}]{herbigdahm} Herbig G. H., Dahm S. E., 2006, AJ, 131, 1530. 
\bibitem[\protect\citeauthoryear{Herbig, Petrov \& Duemmler}{Herbig et al.}{2003}]{herbigetal} Herbig G. H., Petrov P. P., Duemmler R., 2003, ApJ, 595, 384
\bibitem[\protect\citeauthoryear{Ioannidis \& Froebrich}{2012}]{ioannidisfroebrich} Ioannidis G., Froebrich D., 2012, MNRAS, 425, 1380
\bibitem[\protect\citeauthoryear{Jones \& Herbig}{1982}]{jonesherbig} Jones B. F., Herbig G. H., 1982, AJ, 87, 1223
\bibitem[\protect\citeauthoryear{Kazarovets, Reipurth \& Samus}{Kazarovets et al.}{2011}]{kazarovetsetal} Kazarovets E. V., Reipurth B., Samus N. N., 2011, Peremennye Zvezdy, 31, No. 2.
\bibitem[\protect\citeauthoryear{Khanzadyan et al.}{2012}]{khanzadyanetal} Khanzadyan T., Davis C. J., Aspin C., Froebrich D., Smith M. D., Magakian T. Yu., Movsessian T., Moriarty-Schieven G. H., Nikogossian E. H., Pyo T.-S., Beck T. L., 2012, A\&A, 542, 111
\bibitem[\protect\citeauthoryear{Landolt}{1992}]{landolt} Landolt A. U., 1992, AJ, 104, 34
\bibitem[\protect\citeauthoryear{Larson}{1980}]{larson} Larson R. B., 1980, MNRAS, 190, 321
\bibitem[\protect\citeauthoryear{Lodato \& Clarke}{2004}]{lodatoclarke} Lodato G., Clarke C. J., 2004, MNRAS, 353, 841
\bibitem[\protect\citeauthoryear{Magakian \& Movsessian}{1995}]{magakianmovsessian1995} Magakian T. Yu., Movsessian T. A., 1995, A\&A, 295, 504
\bibitem[\protect\citeauthoryear{Magakian \& Movsessian}{1997}]{magakianmovsessian} Magakian T. Yu., Movsessian T. A., 1997, in Malbet F., Castets A., eds, Poster proc. IAU Symp. 182, Low Mass Stars Formation - from Infall to Outflow, Observatoire de Grenoble, p.158
\bibitem[\protect\citeauthoryear{Magakian et al.}{2007}]{magakianetal1} Magakian T. Yu., Aspin C., Pyo T.-S., Movsessian T. A., Nikogossian E. H, Smith M. D., Moiseev A., 2007, in Poster proc. IAU Symp. No. 243, Star-Disk Interaction in Young Stars, http://www.iaus243.org/IMG/pdf/Magakian1.pdf
\bibitem[\protect\citeauthoryear{Magakian et al.}{2010}]{magakianetal2010} Magakian T. Yu., Nikogossian E. H., Aspin C., Pyo T.-S., Khanzadyan T., Movsessian T. A., Smith M. D., Mitchison S., Davis Ch. J., Beck T. L., Moriarty-Schieven G. H., AJ, 139, 969, 2010
\bibitem[\protect\citeauthoryear{Meyer \& Calvet}{1997}]{meyercalvet} Meyer M. R., Calvet N. 1997, AJ, 114, 288
\bibitem[\protect\citeauthoryear{Mickaelian, Sarkissian \& Sinamyan}{Mickaelian et al.}{2012}]{mickaelianetal} Mickaelian A. M., Sarkissian A., Sinamyan P. K., 2012, in Griffin R. E. M., Hanisch R. J. and Seaman R., eds, Proc. IAU Symp, No. 285, New Horizons in Time-Domain Astronomy, p. 366
\bibitem[\protect\citeauthoryear{Miller et al.}{2011}]{milleretal} Miller A. A., Hillenbrand L. A., Covey K. R. and 28 co-authors, 2011, ApJ, 730, 80
\bibitem[\protect\citeauthoryear{Miyazaki et al.}{2002}]{miyazakietal}Miyazaki S., Komiyama Y., Sekiguchi M. and 12 co-authors, 2002, PASJ, 54, 833
\bibitem[\protect\citeauthoryear{Moiseev \& Egorov}{2008}]{moiseevegorov} Moiseev A. V., Egorov O. V., 2008, Astroph. Bull., 63, 181 
\bibitem[\protect\citeauthoryear{Monet et al.}{2003}]{monetetal} Monet D. G., Levine S. E., Canzian B. and 26 co-authors, 2003, AJ, 125, 984
\bibitem[\protect\citeauthoryear{Movsessian \& Magakian}{2004}]{movsessianmagakian} Movsessian T. A., Magakian T. Yu., 2004, Astronomy Rep., 48, 988
\bibitem[\protect\citeauthoryear{Movsessian et al.}{2000}]{movsessianetal}Movsessian T., Boulesteix J., Gach J.-L., Zaratsian S., 2000, Baltic Astron., 9, 652
\bibitem[\protect\citeauthoryear{Movsessian et al.}{2003}]{movsessianetal1} Movsessian T. A., Khanzadyan T., Magakian T. Yu., Smith M. D., Nikogossian E., 2003, A\&A, 412, 147
\bibitem[\protect\citeauthoryear{Movsessian et al.}{2006}]{movsessianetal2} Movsessian T. A., Khanzadyan T., Aspin C., Magakian T. Yu., Beck T., Moiseev A., 2006, A\&A, 455, 1001
\bibitem[\protect\citeauthoryear{Movsessian et al.}{2008}]{movsessianetal3} Movsessian T. A., Magakian T. Yu., Sargsyan D. M., Nikogossian E. H., 2008, Astrophysics, 51, 387 
\bibitem[\protect\citeauthoryear{Reipurth}{1989}]{reipurth} Reipurth B. 1989, Nature, 340, 42
\bibitem[\protect\citeauthoryear{Reipurth \& Aspin}{1997}]{reipurthaspin1} Reipurth B., Aspin C., 1997, AJ, 114, 2700
\bibitem[\protect\citeauthoryear{Reipurth \& Aspin}{2004}]{reipurthaspin2} Reipurth B., Aspin C., 2004, ApJ, 608, L65
\bibitem[\protect\citeauthoryear{Reipurth \& Aspin}{2010}]{reipurthaspin3} Reipurth B., Aspin C., 2010, in Harutynian H. et al., eds, Proc. conf. Evolution of Cosmic Objects through their Physical Activity, Yerevan, p.19
\bibitem[\protect\citeauthoryear{Reipurth \& Bally}{2001}]{reipurthbally} Reipurth B., Bally J., 2001, ARA\&A, 39, 403
\bibitem[\protect\citeauthoryear{Reipurth et al.}{1993}]{reipurthetal1993} Reipurth B., Chini R., Kr\"ugel E., Kreysa E., Sievers A., 1993, A\&A, 273, 221
\bibitem[\protect\citeauthoryear{Reipurth et al.}{2002}]{reipurthetal2002} Reipurth B., Hartmann L., Kenyon S. J., Smette A., Bouchet P., 2002, AJ, 124, 2194.
\bibitem[\protect\citeauthoryear{Reipurth et al.}{2007}]{reipurthetal2007} Reipurth B., Aspin C., Beck T., Brogan C., Conneley M. S., Herbig G. H., 2007, AJ, 13, 1000
\bibitem[\protect\citeauthoryear{Rowles \& Froebrich}{2009}]{rowlesfroebrich} Rowles J., Froebrich D., 2009, MNRAS, 395, 1640
\bibitem[\protect\citeauthoryear{Rydgren \& Vrba}{1983}]{rydgrenvrba} Rydgren A. E., Vrba F. J. 1983, AJ, 88, 1017
\bibitem[\protect\citeauthoryear{Scholz, Froebrich \& Wood}{Scholz et al}{2013}]{scholzetal} Scholz A., Froebrich, D., Wood K. 2013, MNRAS, in press 
\bibitem[\protect\citeauthoryear{Stockton, Chesley \& Chesley}{Stockton et al.}{1975}]{stocktonetal} Stockton A., Chesley D., Chesley S., 1975, ApJ, 199, 406


\end{thebibliography}
\end{document}